\documentclass[letterpaper,english,aps,preprint,nofootinbib,superscriptaddress,byrevtex,showpacs]{revtex4}
\usepackage[T1]{fontenc}
\usepackage[latin1]{inputenc}
\usepackage{amsmath}
\usepackage{amssymb}

\makeatletter

\providecommand{\tabularnewline}{\\}

\usepackage{amssymb}

\usepackage{graphicx}
\usepackage{amssymb}
\usepackage{color}
\usepackage{epsfig}
\providecommand{\tabularnewline}{\\}
\usepackage{bm}

\newcommand{\bra}[1]{\langle #1 \vert}
\newcommand{\ket}[1]{\vert #1 \rangle}

\usepackage{ulem}
\usepackage{soul}
\allowdisplaybreaks

\usepackage{babel}

\usepackage{babel}
\makeatother
\begin{document}

\preprint{LA-UR-05-9140}

\title{Parity Nonconservation in Elastic $\vec{p}\, p$ Scattering}

\author{C.-P. Liu}

\email{liu@kvi.nl}

\affiliation{KVI, Zernikelaan 25, Groningen 9747 AA, The Netherlands}

\thanks{Present address: T-16, Theoretical Division, Los Alamos National
Laboratory, Los Alamos, NM 87545}

\author{C. H. Hyun}

\email{hch@meson.skku.ac.kr}

\affiliation{Department of Physics, Seoul National University, Seoul, 151-742,
Korea}

\affiliation{Institute of Basic Science, Sungkyunkwan University, Suwon 440-746,
Korea}

\author{B. Desplanques}

\email{desplanq@lpsc.in2p3.fr}

\affiliation{Laboratoire de Physique Subatomique et de Cosmologie (UMR CNRS/IN2P3-UJF-INPG),\\
 F-38026 Grenoble Cedex, France}

\begin{abstract}
By looking at the parity-nonconserving (PNC) asymmetries at different
energies in $\vec{p}\, p$ scattering, it is in principle possible
to determine the PNC $\rho NN$ and $\omega NN$ couplings of a single-meson-exchange
model of the PNC $NN$ force. The analysis of the experimental data
at 13.6, 45 and 221 MeV simultaneously has been performed by Carlson
\textit{et al.}, who concluded to an agreement with the original DDH
estimates for the PNC meson-nucleon couplings. In this work, it is
shown first that the comparison with updated hadronic predictions
of these couplings rather suggests the existence of some discrepancy
for the PNC $\omega NN$ coupling. The effect of variations on the
strong coupling constants and introduction of cutoffs in the one-boson-exchange
weak potential is then investigated. As expected, it turns out that
the resulting asymmetry is quite sensitive to these parameters regardless
of the energy. However the above discrepancy persists. The dependence
of this conclusion on various ingredients entering an improved description
of the PNC $NN$ force is also examined. These include the two-pion
resonance nature of the rho meson and some momentum dependence of
the isoscalar PNC $\rho NN$ vertex. It is found that none of these
corrections is able to remove or even alleviate the above discrepancy.
Their impacts on the theoretical determination of the vector meson-nucleon
couplings, the description of the PNC force in terms of single-meson
exchanges, or the interpretation of measurements, are finally examined. 
\end{abstract}

\pacs{24.80.+y, 21.30.-x, 25.40.Cm}

\date{\today{}}

\maketitle

\section{Introduction \label{sec:intro}}

It has been proposed that measurements of the parity-nonconserving
(PNC) longitudinal asymmetry in $\vec{p}\, p$ scattering at different
energies could provide a way to disentangle the separate contributions
to the PNC $NN$ force due to $\rho$- and $\omega$-meson exchanges~\cite{sim-cjp88}.
For some time, accurate measurements were available only at the low
energies of 13.6~MeV~\cite{ever-plb91} and 45~MeV~\cite{kist-prl87}.
Not until recently that a measurement, less accurate though, has been
finished at the higher energy of 221~MeV~\cite{berdoz-prl01} which
makes the above analysis possible. This task has been done by Carlson
\textit{et al.}~\cite{csbg-prc02}, who claimed that the results
so obtained do not disagree with the largest range estimated by Desplanques,
Donoghue and Holstein (DDH)~\cite{ddh80}. A rough understanding
of the measurements is as follows. At the highest energy point (221~MeV),
where the contribution of the $S$ to $P$ $NN$ states vanishes,
the dominant contribution comes from the $P$ to $D$ transition.
It turns out that the corresponding $\omega$-meson-exchange contribution
is suppressed. As a result, this point allows one to fix the $\rho NN$
coupling, $h_{\rho}^{pp}$. Looking now at the low-energy points,
it is found that the $\rho$-exchange force so derived generates PNC
asymmetries larger than the measured ones. Accounting for the experiments
is obtained by fitting the other part of the force due to an $\omega$-meson
exchange, which fixes the $\omega NN$ coupling, $h_{\omega}^{pp}$.
In absence of experimental error, $h_{\omega}^{pp}$ appears to have
a positive sign, opposite to the negative one of the DDH {}``best-guess''
value, and a size at the extreme limit of the estimated range. 

Besides the DDH work, there are many predictions for PNC meson-nucleon
couplings in the literature. Most of them correspond to contributions
already included in the DDH work (see Ref.~\cite{ber98} for references).
Two updated ranges for the PNC couplings are given in Refs.~\cite{des-npa80,fcdh-prc91},
and they do not leave much room for a positive value of the $\omega NN$
coupling, either. This makes it more difficult to accommodate the
value derived from the analysis by Carlson \textit{et al.}~\cite{csbg-prc02}.
Moreover, as especially noticed by Feldman \textit{et al.}~\cite{fcdh-prc91},
predictions for $\rho$ and $\omega$ couplings are not independent
of each other. According to this observation, a larger $h_{\omega}^{pp}$
would thus imply a $h_{\rho}^{pp}$ algebraically larger than the
DDH {}``best-guess'' value, rather than smaller as found in the
analysis by Carlson \textit{et al.}~\cite{csbg-prc02}. An approach
quite different from the DDH one was taken by Kaiser and Meissner~\cite{km-npa},
which used the chiral-soliton model. Their predictions differ from
the {}``best-guess'' values but nevertheless fit into the estimated
range. Actually, they could be approximately obtained from the DDH
work by weighting differently the various contributions considered
there and taking into account the specific dependence of the coupling
constants on the meson squared momentum, $q^{2}$. While DDH estimates
are in principle made at the meson mass ($q^{2}=m^{2}$), Kaiser and
Meissner's ones are given at $q^{2}=0$. The momentum dependence,
which was accounted for very roughly in the DDH work, has been looked
at in detail later on by Kaiser and Meissner in their framework \cite{km2-npa}.
It is found to be especially important for the isoscalar PNC $\rho NN$
coupling. 

>From looking at the different hadronic predictions, it appears very
unlikely that $h_{\omega}^{pp}$ can acquire a positive value. But,
before jumping to the speculation of what could go wrong in these
hadronic calculations, it is important to check the analysis which
depends, in fact quite sensitively, on various issues in the two-nucleon
dynamics. In the past, a lot of these issues have been surveyed by
Simonius~\cite{sim-cjp88}, Nessi-Tedaldi and Simonius~\cite{nsim-plb88},
Driscoll and Miller~\cite{dm-prc89a,dm-prc89b}, and Carlson \textit{et
al.}~\cite{csbg-prc02}. The aim of this current work is to study
if there is some missing two-nucleon dynamics, besides what has been
considered before, which could possibly restore $h_{\omega}^{pp}$
to more conventional values anticipated by existing hadronic calculations. 

On the basis of the DDH {}``best-guess'' values of meson-nucleon
couplings, it is generally considered that the contribution to PNC
effects in $\vec{p}\, p$ scattering is dominated by the $\rho$-meson
exchange. Although the contributions from the $\omega$-meson exchange
are not negligible at low-energy data points, they only constitute
about a 20\% or $-30$\% correction, based on DDH \char`\"{}best-guess\char`\"{}
values or the fitted values by Carlson \textit{et al.}, respectively.
It is therefore appropriate to concentrate on the $\rho$-meson-exchange
contribution at a first step. Taking into account the uncertainty
of the $\rho NN$ coupling, one can temporarily fix $h_{\rho}^{pp}$
to reproduce the low-energy measurements, which are also the most
accurate ones. When this is done, it is found that the measured asymmetry
at the highest energy point (221 MeV) is missed by a factor of about
2. Therefore, any effect that could enhance the transition from $P$
to $D$ states (dominant for 221 MeV) with respect to the one from
$S$ to $P$ states (dominant for 13.6 and 45 MeV) is of relevance
for our purpose. Possibilities can be: (1) a larger vector-meson tensor
coupling $\kappa_{V}$~\cite{hp-npb75}, (2) hadronic form factors
at the strong-interaction vertex, (3) the two-pion resonance nature
of the rho meson~\cite{cb-npb74} and (4) the momentum dependence
of the weak meson-nucleon vertex~\cite{km2-npa,ber98}. For the last
three cases, the enhancement can be naively expected from the resulting
longer range of the PNC $NN$ force, which generally favors transition
amplitudes involving higher orbital angular momenta. However, it should
be noted that the first and third cases may not be independent~\cite{hp-npb75}.
For the parity-conserving (PC) $NN$ force, we use the AV18 model~\cite{wss-pr95}. 

This paper is structured as follows. In Sec. II, the definition of
the PNC longitudinal asymmetry and its analytic form are given. In
Sec. III, we concentrate on the description of the PNC vector-meson-exchange
potential, especially for the $\rho$-meson part. This involves standard
variations of this potential but also less-known ones. We show in
detail how the standard meson-exchange potential is extended to incorporate
the $2\pi$-exchange contribution and the form factor of the PNC vertex.
The asymmetries resulted from different variations of two-nucleon
dynamics are presented in Sec. IV. Their implications are discussed
and new values of weak couplings are obtained from a least-$\chi^{2}$
fit to the measurements. The conclusion follows in Sec. V.

\section{Basic Formalism}

\label{sec:formalism}

The longitudinal asymmetry for nucleon scattering, with an incident
energy $E$ and a scattering angle $\theta$, is defined as\begin{equation}
A_{L}(E,\theta)=\frac{\sigma_{+}(E,\theta)-\sigma_{-}(E,\theta)}{\sigma_{+}(E,\theta)+\sigma_{-}(E,\theta)}\,,\end{equation}
 where $\sigma_{+}$ and $\sigma_{-}$ are differential cross sections
for projectiles of positive and negative helicities, respectively.
In theoretical analyses, however, it is the so-called {}``nuclear''
total asymmetry, $A_{L}^{tot}(E)$, that is often used~\cite{sim-plb72,sim-npa74,sim-cjp88,nsim-plb88,dm-prc89a,csbg-prc02}.
For processes involving Coulomb interactions, such as $\vec{p}\, p$
scattering in this discussion, the total asymmetry is in fact ill-defined,
because total cross sections diverge. The remedy is to remove the
pure Coulomb contribution from the total cross section: by the optical
theorem, the total cross section can be related to the forward scattering
amplitude $\bar{f}(E,\theta=0)$ as\begin{equation}
\sigma^{tot}=\frac{4\,\pi}{k}\,\textrm{Im}[\bar{f}(E,\theta=0)]\,,\end{equation}
 where $k$ is the relative momentum. One can then subtract the pure
Coulomb scattering amplitude $f_{C}$, which is singular at $\theta=0$,
and use the remaining regular {}``nuclear'' scattering amplitude,
$f=\bar{f}-f_{C}$, to define $A_{L}^{tot}(E)$. 

After the spin sums are carried out, the {}``nuclear'' total asymmetry
for $\vec{p}\, p$ scattering takes the following form 

\begin{equation}
A_{L}^{tot}(E)=\frac{\textrm{Im}\left[\tilde{f}_{10,00}(E,0)+\tilde{f}_{00,10}(E,0)\right]}{\textrm{Im}\left[{\displaystyle {\sum_{S,M_{S}}}}f_{SM_{S},SM_{S}}(E,0)\right]}\,,\end{equation}
 where the subscripts $S'M'_{S},SM_{S}$ denote the final and initial
two-body spin states, respectively. The notation $\tilde{f}$ is used
to remind a PNC scattering amplitude: in order to maintain the Pauli
principle for a $p\, p$ system, a spin change must be accompanied
by an orbital angular momentum change, that is, a parity change, too. 

In this work, we treat the PNC interaction, $V_{PNC}$, as a perturbation.
The unperturbed wave functions are solved numerically from the Lippmann-Schwinger
equation\begin{equation}
\ket{\psi}^{(\pm)}=\ket{\phi}^{(\pm)}+\frac{1}{E-H_{0}-V_{C}\pm i\,\epsilon}\, V_{S}\,\ket{\psi}^{(\pm)}\,,\end{equation}
 where $V_{C}$ and $V_{S}$ are the Coulomb and strong interactions,
respectively; and $\ket{\phi}^{(\pm)}$ is the solution of Coulomb
scattering. 

The PC scattering amplitude is given by the following formula\begin{eqnarray}
f_{S'M'_{S},SM_{S}}(E,\theta) & = & \sqrt{4\,\pi}\,\sum_{JLL'}\sqrt{2L+1}\,\epsilon_{L'S'}\,\epsilon_{LS}\,\langle L'(M_{S}-M'_{S}),S'M'_{S}|JM_{S}\rangle\nonumber \\
 &  & \times\langle L0,SM_{S}|JM_{S}\rangle\, Y_{L'(M_{S}-M'_{S})}(\theta)\,{\textrm{e}}^{i\sigma_{L'}}\,\frac{S_{L'S',LS}^{J}(k)-\delta_{L',L}\delta_{S',S}}{i\, k}\,{\textrm{e}}^{i\sigma_{L}},\end{eqnarray}
 where $\epsilon_{LS}$ enforces the Pauli principle: $L+S$ has to
be even; $\sigma_{L}$ is the Coulomb phase shift for the $L$-wave,
and the $S$-matrix element $S_{L'S',LS}^{J}$ can be determined from
the corresponding {}``nuclear'' partial-wave phase shifts. 

The PNC scattering amplitude is calculated by the distorted-wave Born
approximation (DWBA) \begin{equation}
\tilde{f}_{S'M'_{S},SM_{S}}(E,\theta)=-\frac{\mu}{2\,\pi}\,^{(\mp)}\bra{\vec{k}',S'M'_{S}}V_{PNC}\ket{k\,\hat{z},SM_{S}}^{(\pm)}\,,\end{equation}
 where $|\vec{k}'|=|\vec{k}|$, $\hat{k}'\cdot\hat{k}=\cos\theta$,
and $\mu=m_{p}/2$ is the reduced mass. Comparing two recent works,
Refs.~\cite{dm-prc89a} and~\cite{csbg-prc02}, with the former
using DWBA and the latter being numerically exact, treating $V_{PNC}$
as a first-order perturbation is a well-justified approximation. 

It should be noted that $A_{L}^{tot}$, though being well-defined
and easily calculable, is not a quantity which an experiment directly
measures. There are two major setups for $\vec{p}\, p$ scattering:
the scattering-type (for low-energy protons like Refs.~\cite{ever-plb91,kist-prl87})
and the transmission-type (for high-energy protons like Ref.~\cite{berdoz-prl01})
experiments. The former one measures the weighted, average asymmetry
within a selected angular range $[\theta_{1},\theta_{2}]$. The latter
measures the total asymmetry greater than a critical angle $\theta_{c}$,
as the beam in the angular range of $[0,\theta_{c}]$ is extracted
to analyze the transmission rate so that the total cross section between
$[\theta_{c},\pi]$ can be inferred. Since none of these experiments
has full angular coverage and is able to turn off the Coulomb interaction,
some theoretical correction is needed when converting an experimental
asymmetry to $A_{L}^{tot}$. For these issues, we refer readers to
Refs.~\cite{dm-prc89a,csbg-prc02} and publications of individual
experiments for more details.

\section{The vector-meson exchange potential}

\label{sec:pots} The one-meson-exchange PNC $NN$ potential, often
used in the literature, refers to the expression given in Ref.~\cite{ddh80}.
For a $p\, p$ ($n\, n$) system, where only $\rho$ and $\omega$
mesons contribute, this potential can be generalized to the following
form \begin{eqnarray}
V_{PNC}^{pp\,(nn)}(\bm{r}) & = & -\frac{g_{\rho NN}}{m_{N}}\,\Bigg(h_{\rho}^{0}\,\,\bm{\tau}_{1}\cdot\bm{\tau}_{2}+\frac{h_{\rho}^{1}}{2}\;(\tau_{1}^{z}+\tau_{2}^{z})+\frac{h_{\rho}^{2}}{2\sqrt{6}}\,\,(3\tau_{1}^{z}\tau_{2}^{z}-\bm{\tau}_{1}\cdot\bm{\tau}_{2})\Bigg)\nonumber \\
 &  & \hspace{1cm}\times\Big((\bm{\sigma}_{1}-\bm{\sigma}_{2})\cdot\{\bm{p},\, f_{\rho+}(r)\}-(\bm{\sigma}_{1}\times\bm{\sigma}_{2})\cdot\,\hat{\bm r}\, f_{\rho-}(r)\Big)\nonumber \\
 &  & -\frac{g_{\omega NN}}{m_{N}}\,\Bigg(h_{\omega}^{0}+\frac{h_{\omega}^{1}}{2}\;(\tau_{1}^{z}+\tau_{2}^{z})\Bigg)\nonumber \\
 &  & \hspace{1cm}\times\Big((\bm{\sigma}_{1}-\bm{\sigma}_{2})\cdot\{\bm{p},\, f_{\omega+}(r)\}-(\bm{\sigma}_{1}\times\bm{\sigma}_{2})\cdot\,\hat{\bm r}f_{\omega-}(r)\Big),\label{eq:modf1}\end{eqnarray}
 where $m_{N}$ represents the nucleon mass, and $g_{xNN}$'s and
$h_{x}^{i}$'s denote respectively the strong and the weak meson-nucleon
coupling constants for the meson $x$ and isospin $i$.%
\footnote{We notice that the strong and weak couplings are phase dependent.
The convention retained here, usually employed in the field, corresponds
to positive values of the former ones when the $f_{x\pm}(r)$ functions
are given by the Yukawa-like functions given in Eqs. (\ref{eq:fp-yuk},
\ref{eq:fm-yuk}).%
}

The radial functions $f_{x\pm}(r)$ contain important information
about the meson-exchange mechanism such as its range and vertex form
factor, and will be the main variable to be studied in this work.
In the original DDH model, where a point-like ({}``bare'') meson-nucleon
vertex is assumed, they are simply related to the Yukawa function
$f_{x}(r)$ as\begin{align}
f_{x+}^{\textrm{bare}}(r) & =f_{x}(r)\equiv\frac{\textrm{e}^{-m_{x}\, r}}{4\,\pi\, r}\,,\label{eq:fp-yuk}\\
\hat{\bm r}\, f_{x-}^{\textrm{bare}}(r) & =-i\,(1+\kappa_{x})\,[\bm p\,,\, f_{x}(r)]\,,\label{eq:fm-yuk}\end{align}
 where $\kappa_{x}$ is the strong tensor meson-nucleon coupling with
$\kappa_{\rho,\omega}=\kappa_{V,S}$ ($V$ for isovector and $S$
for isoscalar) respectively. Modifications of the above standard PNC
potential to be considered in this work include: (1) variations of
the tensor coupling $\kappa_{V}$ and the introduction of cutoff form
factors at the meson-nucleon vertices, (2) the description of the
$\rho$-meson as a two-pion resonance, and (3) specific PNC meson-nucleon
vertices. All these changes involve different forms of $f_{x\pm}(r)$
which will be precised in the following subsections. We also note
that $f_{x\pm}(r)$ can have isospin dependence -- though it is not
manifest in Eq. (\ref{eq:modf1}) -- and a superscript denoting the
isospin will be added whenever more clarification is necessary. 

When $f_{x\pm}(r)$ does not have isospin dependence, the isospin
matrix elements can be easily evaluated and this gives rise to a $V_{PNC}^{pp}$
depending on two combinations of the weak-coupling constants\begin{align}
h_{\rho}^{pp} & =h_{\rho}^{0}+h_{\rho}^{1}+h_{\rho}^{2}/\sqrt{6}\,,\\
h_{\omega}^{pp} & =h_{\omega}^{0}+h_{\omega}^{1}\,,\end{align}
 and their numbers to be used in our analysis are given in Tab.~\ref{tab:weakconstant}.
The set denoted by DDH corresponds to the DDH {}``best-guess'' values~\cite{ddh80}.
As is known, it roughly accounts for the PNC asymmetries measured
at low energy (13.6 and 45 MeV). It could miss however the high-energy
asymmetry at 221 MeV as reminded in the introduction (see detailed
results in Sec.~\ref{subsec:ctsnum} and Tab.~\ref{tab:result-coupling}).
The other set, {}``adj.'', was fitted by Carlson \textit{et al.}~\cite{csbg-prc02}
to the experimental values of $A_{L}$ at the three above energies.
Since then, it has been used to make predictions for PNC effects in
the $n\, p$ system~\cite{scp-prc04}, showing in some cases significant
differences from the DDH {}``best-guess'' predictions, especially
for the PNC mixing parameter relative to the $^{1}S_{0}-\,\!^{3}P_{0}$
transition, $\epsilon^{0}$. 

\begin{table}
\begin{center}\begin{tabular}{ccc}
\hline 
&
 $h_{\rho}^{pp}\,$&
 $h_{\omega}^{pp}\,$\tabularnewline
\hline
DDH \cite{ddh80}&
 $-15.5$&
 $-3.04$\tabularnewline
adj. \cite{csbg-prc02}&
 $-22.3$&
 $+5.17$ \tabularnewline
\hline
\end{tabular}\end{center}

\caption{Weak coupling constants in units of $10^{-7}$.}

\label{tab:weakconstant}
\end{table}

\subsection{Strong coupling constants and monopole form factors~\label{sub:strong-parameters}}

In the analysis by Carlson \textit{et al.}~\cite{csbg-prc02}, while
various modern strong potentials were used to examine the model dependence,
the strong coupling constants, $g_{\rho NN}$, $g_{\omega NN}$, $\kappa_{V}$,
and $\kappa_{S}$, and meson-exchange dynamics were fixed to the CD-Bonn
model~\cite{mac-prc01}. The introduction of monopole form factors
at both the strong and weak meson-nucleon vertices -- to be consistent
with the Bonn model -- results in a modified radial function in $V_{PNC}^{pp}$

\begin{align}
f_{x+}^{\textrm{mono}}(r) & =\frac{\textrm{e}^{-m_{x}\, r}}{4\,\pi\, r}-\frac{\textrm{e}^{-\Lambda_{x}\, r}}{4\,\pi\, r}\,\left[1+\frac{1}{2}\,\Lambda_{x}\, r\left(1-\frac{m_{x}^{2}}{\Lambda_{x}^{2}}\right)\right]\,,\label{eq:fp-mono}\\
\hat{\bm r}\, f_{x-}^{\textrm{mono}}(r) & =-i\,(1+\kappa_{x})\,[\bm p\,,\, f_{x}^{\textrm{mono}}(r)]\,,\label{eq:fm-mono}\end{align}
 where $\Lambda_{x}$ is the momentum cutoff for the $x$-meson exchange.
The values of these parameters are given in the row S4 of Tab.~\ref{tab:strongconstant}.
\begin{table}
\begin{center}\begin{tabular}{ccccccc}
\hline 
&
 $g_{\rho NN}$&
 $g_{\omega NN}$&
 $\kappa_{V}$&
 $\kappa_{S}$&
 $\Lambda_{\rho}$&
 $\Lambda_{\omega}$\tabularnewline
\hline
S1 &
 2.79 &
 8.37 &
 3.70 &
 $-0.12$&
 - &
 - \tabularnewline
S2 &
 2.79 &
 8.37 &
 6.10 &
 0 &
 - &
 - \tabularnewline
S3 &
 2.79 &
 8.37 &
 3.70 &
 $-0.12$&
 1.31 &
 1.50 \tabularnewline
S4 &
 3.25 &
 15.58 &
 6.10 &
 0 &
 1.31 &
 1.50  \tabularnewline
\hline
\end{tabular}\end{center}

\caption{Sets of the strong coupling constants. The cutoffs $\Lambda_{\rho}$
and $\Lambda_{\omega}$ are in units of GeV.}

\label{tab:strongconstant}
\end{table}

In order to explore the role of the strong coupling constants and
the cutoff values, we consider three additional sets, denoted by S1,
S2 and S3 in Tab.~\ref{tab:strongconstant}. The set S1 corresponds
to strong coupling constants we have been using in our previous works
on PNC problems~\cite{lhd-prc03,lhd-prc04}. It involves values of
$\kappa_{V}=3.7$ and $\kappa_{S}=-0.12$ that are favored by the
vector meson dominance. The set S2 mainly differs from the set S1
by a larger value of $\kappa_{V}=6.1$. This value came from an analysis
of pion-nucleon scattering by H\"{o}hler and Pietarinen~\cite{hp-npb75},
and was adopted in the Bonn model. The set S3 corresponds to a modified
set S1 by introducing the same monopole form factor as Ref.~\cite{csbg-prc02}.
The consideration of sets S1, S2 and S3 is useful in that the comparison
between S1 and S2 gives the dependence on the tensor coupling constants,
and the comparison between S1 and S3 shows the role of the hadronic
form factors. 

Among the different sets considered here, it is not clear at first
sight which one is the most realistic, and comments with this respect
should be done. In absence of information, it seems reasonable to
rely on the parameters fixed by some $NN$ interaction model. However,
they might possibly account for physics different from that one they
are supposed to describe. It has been shown that the large $g_{\omega NN}$
in potential models, like the one in line S4 of Tab.~\ref{tab:strongconstant},
could actually simulate a coherent contribution of a bare-$\omega$
exchange (with a coupling of the size given in the other lines) and
a $\rho\,\pi$ exchange~\cite{dr-npa77}. On the other hand, accounting
for hadronic form factors sounds also reasonable at first, but a dispersion
approach to the derivation of the $NN$ interaction ignores them by
definition. Since form factors imply that the particles have inner
structure, their excitations should be considered for consistency.
As a matter of fact, there are cases that can be worked out where
both effects cancel. This indicates that caution is required in dealing
with form factors. Finally, the term involving the tensor coupling
is expected to be associated with a hadronic form factor that drops
faster than for the other terms, which is most often ignored. 

In the following subsection, we present an improved description of
the $\rho$-exchange contribution. It, in particular, involves the
physics underlying the increase of $\kappa_{V}$ from 3.7 to 6.1,
while providing hadronic form factors (including the faster drop-off
of the form factor associated with $\kappa_{V}$).

\subsection{Two-pion exchange contribution}

\label{subsec:2pi}%
\begin{figure}
\begin{center}\includegraphics[%
  width=10cm]{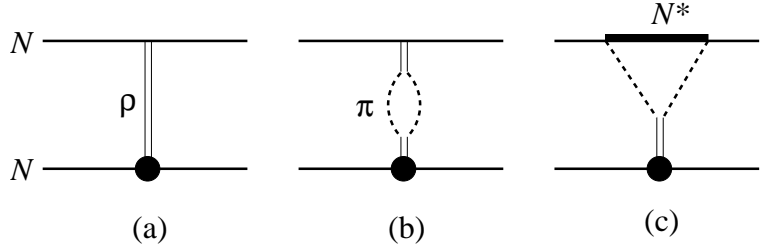}\end{center}

\caption{Graphical representation of the $\rho$ exchange as a stable particle
(a) or taking into account its possible decay into two pions, (b)
and (c). The single solid line denotes a nucleon, the double line
a $\rho$ meson, the dashed line a pion and the thick solid line a
nucleon or a baryon resonance. These last contributions, which involve
intermediate baryons $N$, $\Delta$ and $N^{*}$, are collectively
denoted here by $N^{*}$. The filled circle represents PNC $\rho NN$
vertices.}

\label{fig:twopion}
\end{figure}

In order to account for the two-pion resonance nature of the $\rho$
meson, we follow the work presented in Ref.~\cite{cb-npb74} based
on dispersion relations. In this formalism, only stable particles
are involved and the $\rho$ meson appears indirectly in the transition
amplitude, $N\bar{N}\rightarrow\pi\pi$, through its propagator. To
satisfy unitarity, the width of the $\rho$ meson has to be accounted
for, and this leads to the modification of the free-particle propagator\begin{equation}
\frac{1}{m_{\rho}^{2}-t'}\rightarrow\frac{1}{m_{\rho}^{2}-t'+i\gamma q^{3}(t')}\,,\end{equation}
 where $\gamma$ is related to the $\rho$-meson decay width $\Gamma_{\rho}$
by \begin{equation}
\Gamma_{\rho}=\gamma\, q^{3}(m_{\rho}^{2})\,/m_{\rho}\,,\end{equation}
 $q(t')$ is defined as\begin{equation}
q(t')=\sqrt{\frac{t'}{4}-m_{\pi}^{2}},\end{equation}
 and $t'$ represents the invariant squared mass of the two-pion system
in the $t$-channel of the $NN$ amplitude, on which the integral
in the dispersion relation is performed. The above amplitude has to
be completed for its PC part by a background contribution involving
the exchange in the $t$-channel of the nucleon and the $\Delta$
or $N^{*}$ resonances (collectively denoted as $N^{*}$ in the following,
in absence of ambiguity). The corresponding PNC part is ignored as
it involves new but essentially unknown parameters. 

The next step is to introduce the above $N\bar{N}\rightarrow\pi\pi$
transition amplitudes in the dispersion relation which allows one
to calculate the $NN$ scattering amplitude. In terms of diagrams,
the zero-width $\rho$-meson contribution to the $NN$ interaction,
shown in Fig.~\ref{fig:twopion} (a), is thus replaced by the sum
of contributions depicted in Fig.~\ref{fig:twopion} (b) and in Fig.~\ref{fig:twopion}
(c). For the intermediate baryon states appearing in the last contribution,
we retain, beside the nucleon, the three lowest-lying resonances,
$\Delta(1232)$, $N(1440)$ and $N(1520)$~\cite{cb-npb74}. 

To obtain the potential in configuration space, a standard Fourier
transformation has to be performed. The radial functions $f_{\rho+}(r)$
and $f_{\rho-}(r)$ in Eq. (\ref{eq:modf1}) for the isoscalar, isovector
and isotensor parts now become %
\footnote{The isovector PNC $\rho NN$ coupling was not part of theoretical
frameworks by the time the above work~\cite{cb-npb74} was written.
There is no more reason to ignore it now although the corresponding
contribution is expected to be quite small.%
}

\begin{eqnarray}
f_{\rho+}^{2\pi\,(0,1,2)}(r) & = & \frac{1}{3\,(2\pi)^{3}}\int_{4m_{\pi}^{2}}^{\infty}dt'\,\frac{\textrm{e}^{-r\,\sqrt{t'}}}{r}\,\frac{q^{3}(t')}{\sqrt{t'}}\, g_{\rho+}(t')\,,\label{eq:fplus}\\
f_{\rho-}^{2\pi\,(0,1,2)}(r) & = & \frac{1}{3\,(2\pi)^{3}}\int_{4m_{\pi}^{2}}^{\infty}dt'\,\frac{\textrm{e}^{-r\,\sqrt{t'}}}{r}\left(1+\frac{1}{r\sqrt{t'}}\right)q^{3}(t')\, g_{\rho-}(t')\,.\label{eq:fminus}\end{eqnarray}
 The spectral functions $g_{\rho+}(t')$ and $g_{\rho-}(t')$ are
defined as \begin{eqnarray}
g_{\rho+}(t') & = & \frac{f_{\rho}^{2}}{(m_{\rho}^{2}-t')^{2}+\gamma^{2}q^{6}(t')}+\textrm{Re}\frac{\beta(t')+m_{N}\alpha(t')}{m_{\rho}^{2}-t'+i\gamma q^{3}(t')}\,,\label{eq:gplus}\\
g_{\rho-}(t') & = & \frac{f_{\rho}^{2}(1+\kappa_{V})}{(m_{\rho}^{2}-t')^{2}+\gamma^{2}q^{6}(t')}+\textrm{Re}\frac{\beta(t')}{m_{\rho}^{2}-t'+i\gamma q^{3}(t')}\,,\label{eq:gminus}\end{eqnarray}
 where \begin{equation}
\gamma=\frac{f_{\rho}^{2}}{6\,\pi\; m_{\rho}}\,.\end{equation}
 The quantities, $\alpha(t')$ and $\beta(t')$, appearing in the
above equations are given by \begin{eqnarray}
\alpha(t') & = & \frac{3}{q(t')\,\chi^{2}(t')}\sum_{N^{*}}\left[\frac{G_{N^{*}}^{A}}{q(t')}\left(1-h\;\textrm{tan}^{-1}\frac{1}{h}\right)\right.\nonumber \\
 &  & \left.\hspace*{2.0cm}+\frac{m_{N}}{2\chi(t')}G_{N^{*}}^{B}\left(3h-(1+3h^{2})\,\textrm{tan}^{-1}\frac{1}{h}\right)\right]\,,\label{eq:alpha}\\
\beta(t') & = & -\frac{3}{2q(t')\,\chi(t')}\sum_{N^{*}}G_{N^{*}}^{B}\left[h-(1+h^{2})\,\textrm{tan}^{-1}\frac{1}{h}\right]\,,\label{eq:beta}\end{eqnarray}
 with \begin{eqnarray}
\chi(t') & = & \sqrt{m_{N}^{2}-\frac{t'}{4}}\,,\\
h=h(t') & = & \frac{q^{2}(t')-\chi^{2}(t')+m^{*2}}{2q(t')\,\chi(t')}\,.\end{eqnarray}
 The values of the coefficients $G_{N^{*}}^{A,B}$ used in the present
work are given in Tab.~\ref{tab:nstar}. %
\begin{table}
\begin{center}\begin{tabular}{l|c|c}
\hline 
$N^{*}$&
 $G_{N^{*}}^{A}/(4\pi\, m_{\pi})$&
 $G_{N^{*}}^{B}/(4\pi)$\tabularnewline
\hline
$N$&
 0 &
 14.48 \tabularnewline
$\Delta(1232)$&
 $-21.8-0.97\frac{t'}{m_{\pi}^{2}}$&
 $7.4-0.062\frac{t'}{m_{\pi}^{2}}$\tabularnewline
$N(1440)$&
 $-7.11$&
 2.15\tabularnewline
$N(1520)$&
 $-5.75-0.252\frac{t'}{m_{\pi}^{2}}$&
 $1.26+0.06\frac{t'}{m_{\pi}^{2}}$ \tabularnewline
\hline
\end{tabular}\end{center}

\caption{Coefficients $G_{N^{*}}^{A}$ and $G_{N^{*}}^{B}$ appearing in Eqs.
(\ref{eq:alpha}, \ref{eq:beta}): values for the intermediate baryons
(nucleon and resonances) retained here.}

\label{tab:nstar}
\end{table}

Different values of $f_{\rho}^{2}/(4\pi)$ are referred to in the
literature. They can be related, for instance, to the decay width
of $\rho\rightarrow e^{-}e^{+}$ or to $g_{\rho NN}$ by the hypothesis
of vector meson dominance. In the present estimate, we use the latter.
For consistency with $g_{\rho NN}=2.79$ in the sets S1--S3 (see Tab.~\ref{tab:strongconstant}),
$f_{\rho}^{2}/(4\pi)=2.5$ (2.08 was used in Ref.~\cite{cb-npb74}). 

If one neglects the contribution from intermediate baryons and takes
the limit of $\gamma\rightarrow0$ ($\Gamma_{\rho}$=0) in Eqs.~(\ref{eq:gplus},
\ref{eq:gminus}), the radial functions $f_{\rho\pm}^{2\pi\,(0,1,2)}(r)$
simply reduce to the original Yukawa-like ones $f_{\rho\pm}^{\textrm{bare}}(r)$. 

The two-pion exchange interaction also contains a part with an isovector
character which results from a non-zero $\pi NN$ coupling. Its contribution
to the PNC asymmetry of interest in this work has been calculated
in the past~\cite{sim-npa74}. It has not been considered here however.
While it could represent one half of the low-energy measurements with
the {}``best-guess'' value of this coupling, there are many reasons
to believe that this coupling is actually smaller~\cite{ber98}.
The corresponding contribution is therefore expected to play a minor
role. On the other hand, this contribution looks like a $\rho$-exchange
one~\cite{cb-npb74} and qualitative results obtained here for the
other two-pion exchange contribution, $f_{\rho\pm}^{2\pi\,(0,1,2)}(r)$,
would largely apply to it in any case.

\subsection{Parity-nonconserving $\rho NN$ vertex form factor}

\label{subsec:ff}

Meson-nucleon vertex functions are generally written as the product
of the coupling constant, defined for an on-mass-shell meson ($q^{2}=m^{2}$),
and a form factor which depicts the $q^{2}$ dependence. Among various
empirical choices, the monopole one is often adopted for the strong
vertex, which leads to, \textit{e.g.}, the Bonn potentials~\cite{mhe-pr87,mac-prc01}.
As already mentioned in the subsection~\ref{sub:strong-parameters},
the works of Refs.~\cite{dm-prc89a,csbg-prc02} involve applying
the same monopole form factor also to the weak vertex, which gives
rise to modified radial functions as in Eqs.~(\ref{eq:fp-mono},
\ref{eq:fm-mono}). However, one can certainly speculate about other
possibilities. 

In the DDH work~\cite{ber98}, the coupling constant receives three
contributions: {}``factorization'', {}``parity admixture'' and
{}``sea quarks''. For simplicity, they have been taken as constant.
The two last contributions were estimated by relying on the $SU(6)_{W}$
symmetry and experimental information from non-leptonic hyperon decays
but the authors also worried about symmetry-breaking effects. These
ones could be sizable for the {}``parity-admixture'' contribution
to the isoscalar $\rho NN$ coupling, which can be shown to vanish
at $q^{2}=0$. %
\footnote{This contribution is absent for the isotensor $\rho NN$ coupling
and both isoscalar and isovector $\omega NN$ couplings. The vanishing
of the contribution at $q^{2}=0$ for the isoscalar $\rho NN$ coupling
has some relationship with an anapole moment contribution.%
} This result is due to the cancellation of two contributions with
the same topology but involving intermediate quarks with negative
and positive energies. While the first one is included by using the
$SU(6)_{W}$ symmetry and could be appropriate for an on-mass-shell
meson, the effect of the second one is ignored. To account for the
expected cancellation, the {}``parity-admixture'' contribution to
the isoscalar $\rho NN$ coupling was suppressed by a factor 4 in
getting the {}``best-guess'' values. A refined estimate would suppose
to calculate the $q^{2}$ dependence of the coupling constant, which
was done by Kaiser and Meissner in their framework \cite{km2-npa}.
Although there exists no detailed comparison, their results tend to
support the above analysis. The $q^{2}$ dependence is especially
important for the isoscalar $\rho NN$ coupling. It evidences a feature
which is somewhat unusual for current form factors but is a signature
of the underlying dynamics: a change of sign occurs at $q^{2}=q_{0}^{2}-\bm q^{2}\simeq-m_{\rho}^{2}$. 

We now consider the effect of inserting the above momentum dependence
in the PNC $NN$ interaction. Consistently with the non-relativistic
approach used here, we neglect the energy transfer carried by the
meson and therefore assume $q^{2}\simeq-\bm q^{2}$ in the following.
The isoscalar PNC $\rho NN$ vertex form factor, $\widetilde{F}_{\rho}^{\textrm{KM\,}(0)}(\bm q^{2})$,
in Ref.~\cite{km2-npa} can thus be approximately parametrized as
\begin{eqnarray}
\widetilde{F}_{\rho}^{\textrm{KM}\,(0)}(\bm q^{2})=\left(1-2\frac{\bm{q}^{2}}{\bm{q}^{2}+\Lambda'\,^{2}}\right)\,.\label{eq:modp-km}\end{eqnarray}
 At low-momentum transfer, the parameter $\Lambda'$ has the same
effect as usual cutoff parameters but its role differs at high momentum
transfer (hence a different notation). The sensitivity to this parameter
will be studied in the later section. Assuming the corresponding strong
vertex is still a point-like one, the radial functions for the isoscalar
$\rho$ exchange, modified by Eq. (\ref{eq:modp-km}), now reads \begin{align}
f_{\rho+}^{\textrm{KM}\,(0)}(r) & =\frac{\Lambda'\,^{2}+m_{\rho}^{2}}{\Lambda'\,^{2}-m_{\rho}^{2}}\; f_{\rho}(r)-\frac{2\Lambda'\,^{2}}{\Lambda'\,^{2}-m_{\rho}^{2}}\; f_{\Lambda'}(r)\,,\label{eq:fp-km}\\
f_{\rho-}^{\textrm{KM}\,(0)}(r) & =-i\,(1+\kappa_{\rho})\,[\bm p\,,\, f_{\rho+}^{\textrm{KM}\,(0)}(r)]\,.\label{eq:fm-km}\end{align}

In the limit $\Lambda'\rightarrow\infty$, $f_{\rho+}^{\textrm{KM}\,(0)}(r)$
recovers the standard Yukawa function $f_{\rho}(r)$. In a special
case where $\Lambda'=m_{\rho}$, corresponding to the above mentioned
change of sign at $q^{2}\simeq-m_{\rho}^{2}$, $f_{\rho+}^{\textrm{KM}\,(0)}(r)$
remains finite\begin{eqnarray}
f_{\rho+}^{\textrm{KM}\,(0)}(r)=\frac{1}{4\pi r}\textrm{e}^{-m_{\rho}r}(m_{\rho}r-1)\,,\label{eq:mod-km-rho}\end{eqnarray}
 despite the presence of the factor $\Lambda'\,^{2}-m_{\rho}^{2}$
in the denominator.

\subsection{Resulting potentials}

\begin{figure}
\begin{center}\includegraphics[%
  width=7cm]{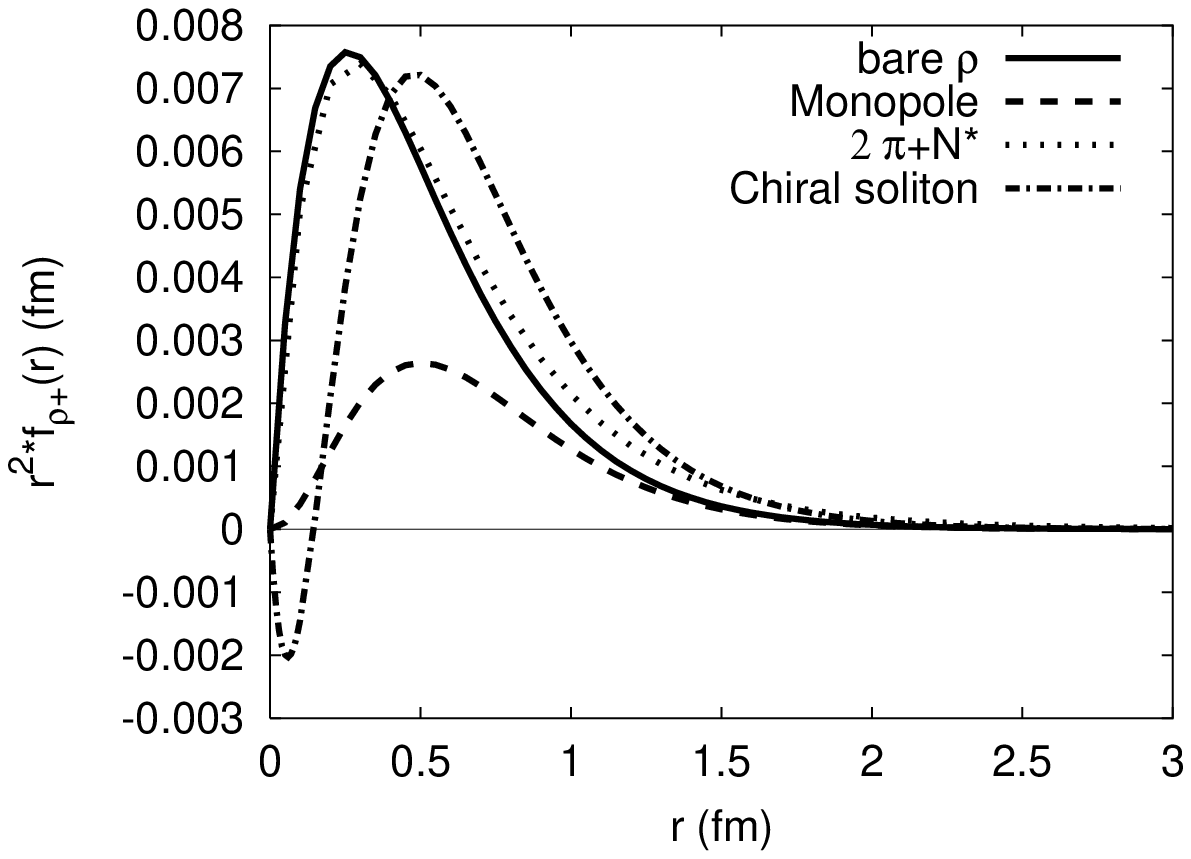}\includegraphics[%
  width=7cm]{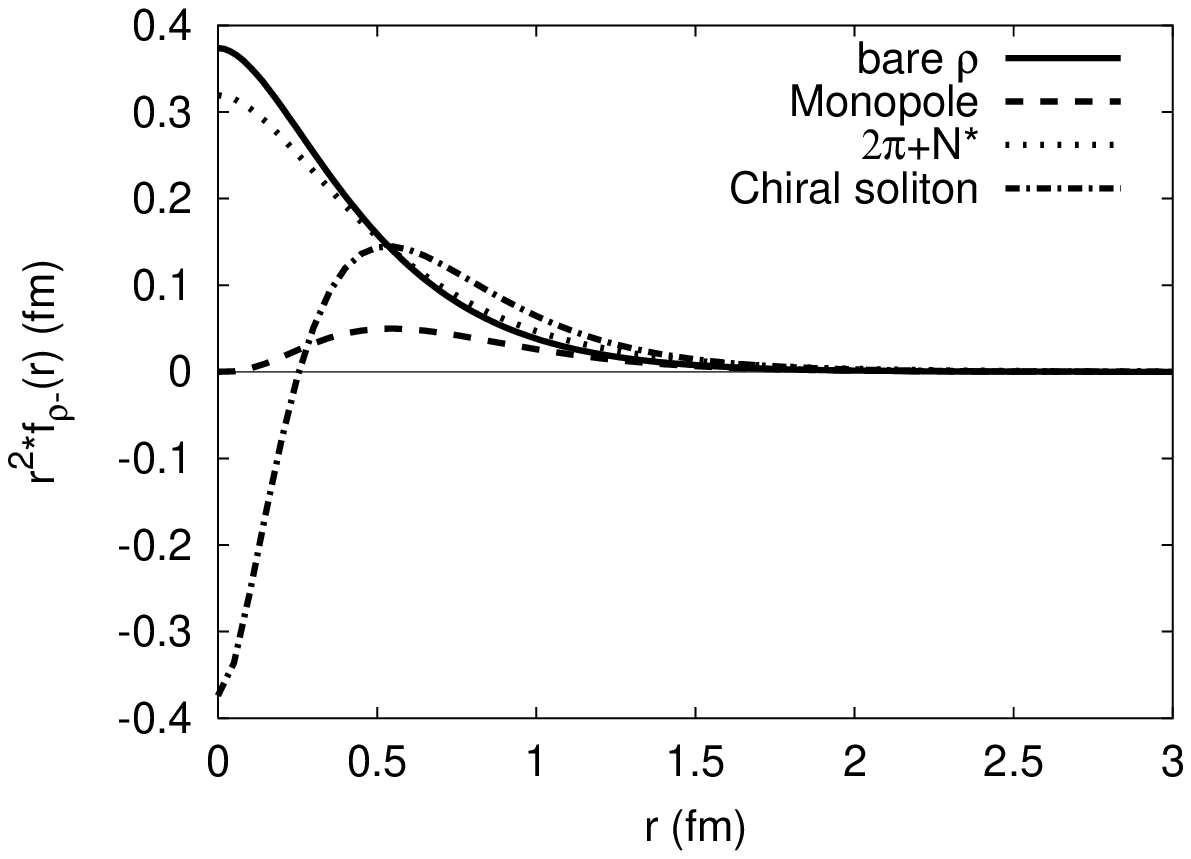}\end{center}

\caption{Yukawa function (bare-$\rho$, continuous) and modified ones due
to form factors with a monopole type at both PC and PNC vertices (Monopole,
dashed), to $2\pi$ and $N^{*}$ corrections ({}``$2\pi+N^{*}$'',
dot), and to PNC form factors obtained from the chiral-soliton model
calculation \cite{km2-npa} (Chiral soliton, dot-dashed). The left
panel is for $f_{\rho+}(r)$, and the right one for $f_{\rho-}(r)$.
Expressions of the potentials are given in the text, Eqs. (\ref{eq:fp-yuk},
\ref{eq:fm-yuk}, \ref{eq:fp-mono}, \ref{eq:fm-mono}, \ref{eq:gplus},
\ref{eq:gminus}, \ref{eq:fp-km}, \ref{eq:fm-km}), and parameters
entering the monopole and chiral-soliton ones, $\Lambda$ and $\Lambda'$,
are given the unique value 1.31 GeV.}

\label{fig:potbare-kmr2}
\end{figure}

The masses of $\pi$, $\rho$ and $\omega$ mesons are set to be 139.0,
771.0 and 783.0 MeV respectively, throughout the calculations. Plotted
in Fig.~\ref{fig:potbare-kmr2} are $f_{\rho+}(r)$ and $f_{\rho-}(r)$
multiplied by $r^{2}$ which appears in the $r$-space integration.
The potentials with the bare-$\rho$ exchange, the monopole form factor,
the $2\pi+N^{*}$ corrections and the PNC chiral-soliton form factor
are represented respectively by solid, dashed, dotted and dot-dashed
lines. 

Compared to the bare-$\rho$-exchange potential, the $2\pi+N^{*}$
corrected ({}``$2\pi+N^{*}$'') one gives a non-negligible enhancement
in the range $0.5\leq r\leq2$ fm. In the remaining regions, however,
these two potentials are almost indistinguishable. For the potentials
with form factors, we show results with $\Lambda'=1.31$ GeV and $\Lambda_{\rho}=1.31$
GeV for the chiral-soliton and monopole ones respectively. Both of
them give significant difference from bare-$\rho$ and {}``$2\pi+N^{*}$''
potentials at $r\leq2$ fm. The chiral-soliton form factor enhances
the potential substantially at $0.4\leq r\leq2$ fm, drops rapidly
at around $r\simeq0.4$ fm and changes sign at $r\leq0.2$ fm. The
change of sign can give a negative contribution to the matrix elements,
but the quantitative estimation is dependent on the shape of the wave
functions. With different $\Lambda'$ values, the shape of the potential
changes. For a value smaller than 1.31 GeV, the change of sign is
shifted to larger $r$ values, and the enhancement in the intermediate
range becomes more significant than for the present potential. On
the other hand, if one increases the $\Lambda'$ value, the potential
becomes more similar to the bare-$\rho$ one. We will show this behavior
explicitly when we discuss the results. Contrary to the PNC chiral-soliton
form factor, a monopole form factor gives suppression in magnitude
over the whole $r$ region. This suppression will give matrix elements
smaller than in the remaining three cases. More importantly, a monopole
form factor makes the potential more sensitive to the cutoff value
than the PNC chiral-soliton form factor is to the value of $\Lambda'$.
We will argue about this point in the forthcoming results. 

Concluding this section, an important observation should be made with
respect to the motivation of the present work. In comparison to the
standard $\rho$-exchange potential, some of the variations we consider
tend to make its range longer. At first sight, the feature which can
possibly enhance the contribution of $P$ to $D$ $NN$ states with
respect to the $S$ to $P$ ones is desirable. This can be checked
by calculating the plane-wave Born amplitude, but a definitive answer
requires a full calculation with distorted wave functions.

\section{Numerical results and discussion}

\label{sec:results}

We here discuss qualitatively the effects of the different variations
on the meson-exchange potential laid down in the previous section.

\subsection{Effect of the coupling constants and monopole form factors}

\label{subsec:ctsnum}%
\begin{table}
\begin{center}\begin{tabular}{c|cccc|c|c}
\hline 
Weak &
\multicolumn{4}{c|}{DDH}&
 adj. &
\tabularnewline
\cline{2-5} 
 Strong &
 S1 &
 S2 &
 S3 &
 S4 &
 S4 &
 Exp.~%
\footnote{These values are taken from Ref.~\cite{csbg-prc02}, assuming the
theoretical corrections have been made.%
} \tabularnewline
\hline
13.6 &
 $-0.96$&
 $-1.33$&
 $-0.66$&
 $-1.13$&
 $-0.92$&
 $-0.95\pm0.15$~\cite{ever-plb91}\tabularnewline
45 &
 $-1.73$&
 $-2.39$&
 $-1.16$&
 $-2.00$&
 $-1.59$&
 $-1.50\pm0.23$~\cite{kist-prl87}\tabularnewline
221 &
 0.43 &
 0.75 &
 0.25 &
 0.52 &
 0.85 &
 $0.84\pm0.29$~\cite{berdoz-prl01} \tabularnewline
\hline
\end{tabular}\end{center}

\caption{Sensitivity of the PNC asymmetry, $A_{L}(\times10^{7})$, to different
choices of weak and strong coupling constants, or to monopole form
factors (see Tabs.~\ref{tab:weakconstant} and \ref{tab:strongconstant}
for their values) and comparison with experiment. }

\label{tab:result-coupling}
\end{table}

In Tab.~\ref{tab:result-coupling}, the results with various chosen
parameter sets (see Tabs.~\ref{tab:weakconstant} and \ref{tab:strongconstant}
for their values) are presented. The effect of the coupling constants
is straightforward: larger coupling constants give larger asymmetries.
As shown in Ref.~\cite{sim-cjp88} that the $S-P$ transition dominates
at the low energies, while the $P-D$ transition does at the high
energies, the $S-P$ transition amplitude is approximately proportional
to \begin{eqnarray}
h_{\rho}^{pp}\, g_{\rho NN}\,(\kappa_{V}+2)+h_{\omega}^{pp}\, g_{\omega NN}\,(\kappa_{S}+2)\,,\label{eq:roughsp}\end{eqnarray}
 and the $P-D$ transition amplitude to \begin{eqnarray}
h_{\rho}^{pp}\, g_{\rho NN}\,\kappa_{V}+h_{\omega}^{pp}\, g_{\omega NN}\,\kappa_{S}\,.\label{eq:roughpd}\end{eqnarray}

Beginning the discussion with a comparison of the predictions and
measurements, we observe that the S1-set results agree with the low-energy
measurements. At 221 MeV, the result is smaller than the lowest experimental
value by about 22\%. For the set S2, the situation is opposite: the
result at 221 MeV is within the error bar, but those at low energies
are off. Since the asymmetry can be well approximated by Eqs.~(\ref{eq:roughsp},
\ref{eq:roughpd}), which are linearly dependent on the strong coupling
constants $g_{xNN}$ and $\kappa$'s, we conclude that those good
at low energy are not good at 221 MeV, and vice versa. As expected,
the asymmetry is sensitive to the strong coupling constants, but this
is almost irrelevant to the resolution of the problem raised in the
introduction. 

We now consider in more detail the sensitivity to the isovector tensor
coupling $\kappa_{V}$. This can be done by comparing results of sets
S2 and S1, or S4 and S3. Evaluating Eq.~(\ref{eq:roughsp}) with
S2 and S1, we obtain the ratio S2/S1 ($S-P$) $\simeq$ 1.36. This
value is comparable to the ratios of S2/S1 at 13.6 and 45 MeV, 1.39
and 1.38, respectively. Equation~(\ref{eq:roughpd}) gives S2/S1
($P-D$) $\simeq$ 1.68, and this value is close to S2/S1 at 221 MeV,
1.74. In a similar way, we can compare S4 and S3. We have the ratios
S4/S3 ($S-P$) $\simeq$ 1.71 and S4/S3 ($P-D$) $\simeq$ 1.96. Our
calculation gives 1.71 and 1.72 at 13.6 and 45 MeV, respectively,
and 2.10 at 221 MeV. In both cases, it is found that changing $\kappa_{V}$
from 3.7 to 6.1 enhances the prediction for the high-energy point
with respect to the low-energy ones. The effect, which is of the order
of 25\%, goes in the direction we looked for. However, for the set
S1, enhancement due to a larger value of $\kappa_{V}$ is still lacking
to fit the high-energy asymmetry within the experimental error bar. 

The role of monopole form factors can be understood by comparing results
of sets S3 and S1 (or S4 and S2 after correcting for different strong
couplings in this case). The ratios are 0.69, 0.69 and 0.58 at 13.6,
45 and 221 MeV respectively (or 0.73, 0.72 and 0.59). One sees a clear
indication that the effect of monopole form factors is in favor of
$S$ to $P$ transitions, contrary to what we would naively expect
from a longer-range interaction. This point will be further examined
later on after other similar long-range effects are also considered. 

\begin{figure}
\begin{center}\includegraphics[%
  width=7cm]{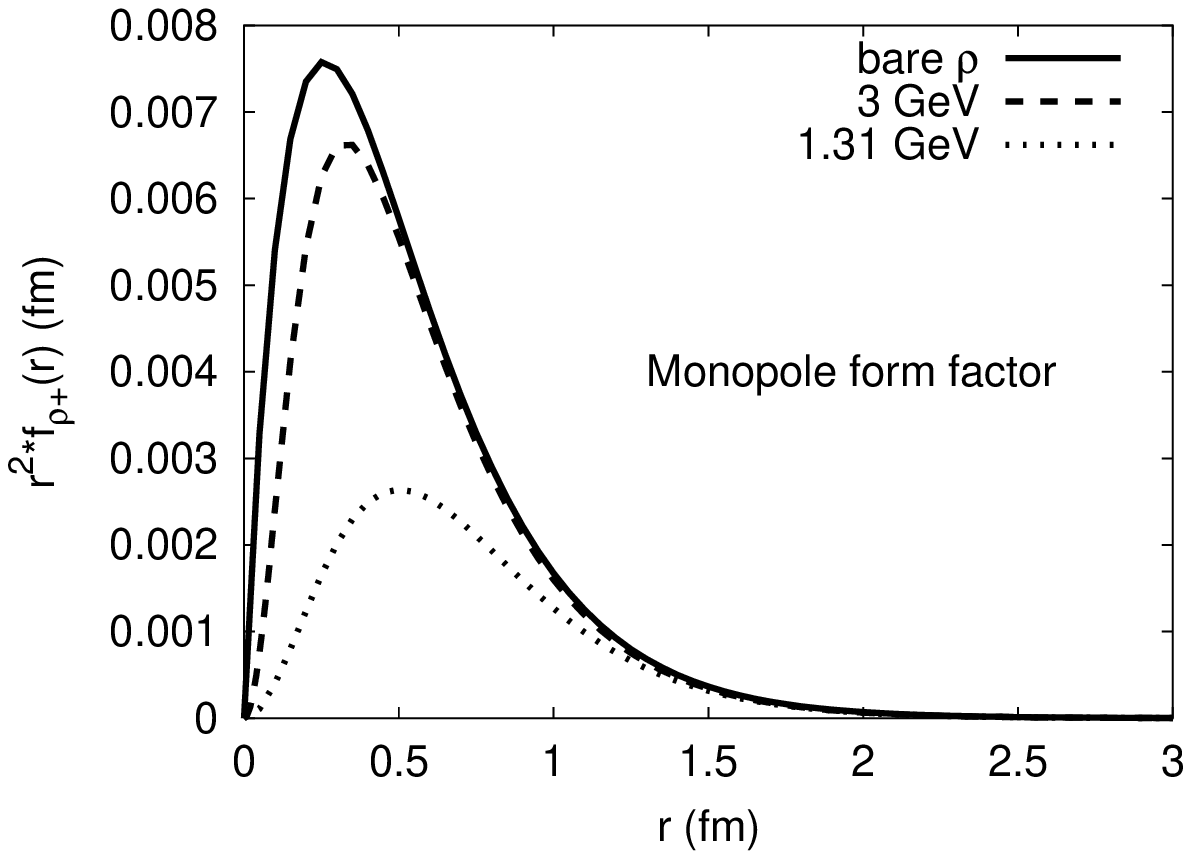}\includegraphics[%
  width=7cm]{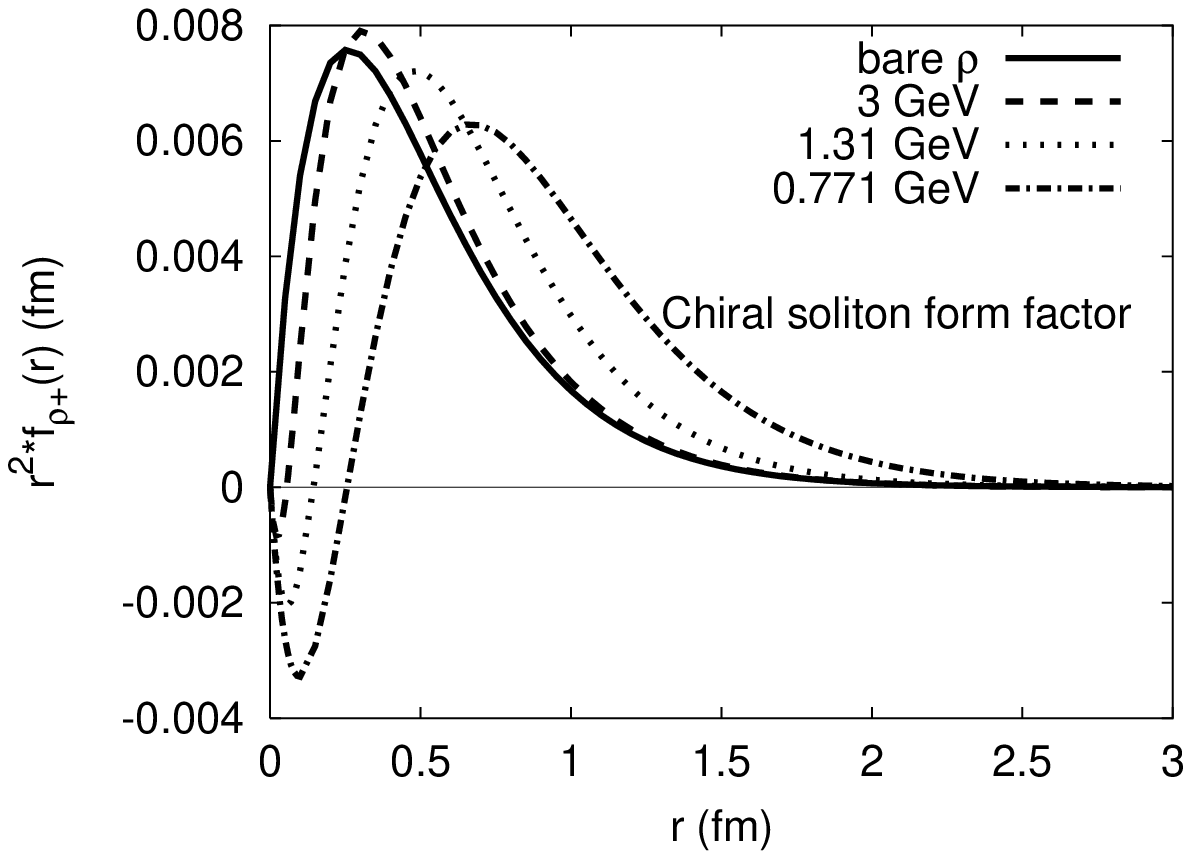}\end{center}

\caption{Modified Yukawa functions multiplied by $r^{2}$: with the square
of monopole form factor (left panel) and the PNC chiral-soliton form
factor (right panel). For illustration, the cutoff $\Lambda$ is given
the values $\infty$ (bare), 3, and 1.31 GeV in one case while the
parameter $\Lambda'$ assumes the values $\infty$ (bare), 3, 1.31
and 0.771 GeV in the other case .}

\label{fig:ff-comp}
\end{figure}
 For a part, the above results can be understood from the behavior
of the potentials. In Fig.~\ref{fig:ff-comp}, the left panel shows
the Yukawa potential modified by monopole form factors, with a multiplication
factor $r^{2}$. The smaller the cutoff value is, the smaller is the
potential for all $r$. Thus, with a smaller cutoff value, the asymmetry
is also smaller in magnitude. The large sensitivity of the potential
to the cutoff value as Fig.~\ref{fig:ff-comp} shows, has its origin
in the expression of the squared monopole form factor in momentum
space, $((\Lambda^{2}-m^{2})/(\Lambda^{2}+\bm{q}^{2}))^{2}$. This
form, which is consistent with the definition of couplings made for
on-mass-shell mesons, implies an overall suppression of PNC amplitudes
at low energy by a factor $((\Lambda^{2}-m^{2})/(\Lambda^{2}))^{2}$
(=0.43 for $\Lambda=1.31$ GeV). This behavior differs from the one
evidenced by other potentials considered below where the factor under
discussion is essentially absent. On the other hand, the difference
between the above suppression factor and the one deduced in the previous
paragraph by comparing S3 and S1 results indicates that the effect
of the potential occurs at distances larger than what the position
of maxima in Fig.~\ref{fig:ff-comp} suggests.

\subsection{Effect of the $2\pi$ and $N^{*}$ corrections~\label{subsec:2pinum} }

The effect of the $2\pi+N^{*}$ corrections is investigated with the
strong parameter sets S1 and S2, and the DDH {}``best-guess'' values
for the weak coupling constants. The results are summarized in Tab.~\ref{tab:result-potential}.
In the column {}``$2\pi+N^{*}$'', the numbers in the parentheses
represent the ratios of results ({}``$2\pi+N^{*}$'')/($\textrm{bare}\,\rho$). 

\begin{table}
\begin{center}\begin{tabular}{c|c|c|c|c|c}
\hline 
&
\multicolumn{2}{c|}{S1}&
\multicolumn{2}{c|}{S2}&
\tabularnewline
\cline{2-3} \cline{4-5} 
&
 bare-$\rho$&
 {}``$2\pi+N^{*}$'' &
 bare-$\rho$&
 {}``$2\pi+N^{*}$'' &
 Exp. \tabularnewline
\hline
13.6 &
 $-0.96$&
 $-1.22$ (1.26) &
 $-1.33$&
 $-1.68$ (1.25) &
 $-0.95\pm0.15$\tabularnewline
45 &
 $-1.73$&
 $-2.11$ (1.21) &
 $-2.39$&
 $-2.92$ (1.22) &
 $-1.50\pm0.23$\tabularnewline
221 &
 0.43 &
 0.52 (1.22) &
 0.75 &
 0.92 (1.22) &
 $0.84\pm0.29$ \tabularnewline
\hline
\end{tabular}\end{center}

\caption{Sensitivity of the PNC asymmetry, $A_{L}\,(\times10^{7})$, to the
effect of the finite $\rho$-width correction of the weak potential.
Weak coupling constants are fixed to the DDH {}``best-guess'' values.
Two sets of strong couplings, S1 and S2, are considered. The numbers
in the parentheses represent the ratios ({}``$2\pi+N^{*}$'')/(bare-$\rho$).}

\label{tab:result-potential}
\end{table}

The {}``$2\pi+N^{*}$'' result evidences a relatively larger enhancement
at 13.6 MeV than at the remaining two energies, but as a whole, the
ratios are similar. For the set S1, the {}``$2\pi+N^{*}$'' potential
increases the asymmetry by 0.26, 0.38 and 0.09 at 13.6, 45 and 221
MeV respectively (in units of $10^{-7}$). Consequently the low-energy
asymmetries exceed the experimental upper limit while the high-energy
one is still below. For the set S2, The amount of increase is larger
than for the set S1: 0.35, 0.53 and 0.17 at 13.6, 45 and 221 MeV respectively
(again in units of $10^{-7}$). Thus, the low-energy predictions,
which are already out of the experimental error bars with the $\rho$-exchange
potential, are further away from experiment. Meanwhile, a relatively
small increase of the high-energy asymmetry keeps the prediction within
error bars. 

To get some insight into the above results, it is interesting to look
at the potentials in Fig.~\ref{fig:potbare-kmr2}. At $r\geq2$ fm,
they show a similar behavior. In comparison to the bare-$\rho$ potential
within the range $0.4\leq r\leq2$ fm, $f_{\rho+}^{2\pi}(r)$ is sizably
larger but $f_{\rho-}^{2\pi}(r)$ is slightly smaller. In the range
$r\leq0.4$ fm, $f_{\rho+}^{2\pi}(r)$ is very similar but $f_{\rho-}^{2\pi}(r)$
is clearly smaller. From the result of the asymmetry in Tab.~\ref{tab:result-potential},
one can deduce that the suppression of $f_{\rho-}^{2\pi}(r)$ at short
distances does not much affect the magnitude of the asymmetry. Therefore,
roughly speaking, a good deal of the difference in the results comes
from the difference of the potentials in the region $0.4<r<2$ fm
and the contribution from $r<0.4$ fm is negligible. 

While the enhancement of the interaction in the range $0.4<r<2$ fm
can explain enhanced asymmetries, an enhancement of asymmetry at the
higher energy point with respect to the low ones, as one would expect
from a longer-range interaction, does not show up. Though the effect
is small ($0\sim3$\%), it is opposite to what could be naively expected.
It is interesting to compare the results with those obtained from
using the plane-wave Born approximation (PWBA). The enhancement of
the asymmetry at low energy would be about 12\% while it reaches 26\%
at high energy (the enhancement for the $P$ to $D$ transition at
low energy is about 60\%). 

As mentioned in Sec.~\ref{sub:strong-parameters}, the enhanced $\kappa_{V}$
value, 6.1, could account for the physics we included here in allowing
for the contribution of nucleon and baryon resonances to the $\pi N$
scattering amplitude (or the $N\bar{N}\rightarrow\pi\pi$ amplitude).
Therefore, results denoted {}``$2\pi+N^{*}$'' with S1 and bare-$\rho$
with S2 in Tab.~\ref{tab:result-potential} should not be independent.
This is supported for a part by the fact that both results deviate
from the bare-$\rho$ S1 contribution by relatively the same amount
for the low-energy measurements (roughly 24\% and 38\%). The difference
is larger for the high-energy point (22\% and 74\%), but this could
be due to the approximate character of treating the effect of extra
contributions to the $\pi N$ scattering amplitude by a constant number.
In principle, the $\kappa_{V}$ contribution is expected to be associated
with a form factor that decreases faster than for the other contributions.

\subsection{Effect of the PNC vertex form factor~\label{subsec:ffnum}}

Results with the monopole form factor have been discussed at the beginning
of the present section (see Tab.~\ref{tab:result-coupling}). We
here consider the effect of the PNC chiral-soliton form factor for
the isoscalar $\rho NN$ coupling. As this form factor could involve
some uncertainty, we also looked at variations of the cutoff parameter
$\Lambda'$ in Eq.~(\ref{eq:modp-km}). Besides the value $\Lambda'=0.771$
GeV, which approximately fits Kaiser and Meissner's estimate~\cite{km2-npa},
we consider the values 1.31 and 3 GeV. The first of these last values
fits the low-momentum dependence of the monopole form factor used
by Carlson \textit{et al.} and the second allows one to make the transition
to the standard point-like $\rho NN$ coupling. The larger $\Lambda$'s
are probably closer to the one inferred from the DDH work, though
quite uncertain. Strong coupling constants are picked up from the
set S1, and DDH {}``best guess'' values are used for the weak coupling
constants. The results with different strong coupling constants, \textit{e.g.,}
S2, can be easily deduced from Eqs.~(\ref{eq:roughsp}, \ref{eq:roughpd}). 

\begin{table}
\begin{center}\begin{tabular}{c|c|c|c|c}
\hline 
$\Lambda'$ (GeV) &
 bare &
 3 &
 1.31 &
 0.771 \tabularnewline
\hline
13.6 &
 $-0.96$&
 $-1.04$&
 $-1.33$&
 $-1.69$\tabularnewline
45 &
 $-1.73$&
 $-1.88$&
 $-2.38$&
 $-2.92$\tabularnewline
221 &
 $0.43$&
 $0.47$&
 $0.61$&
 $0.67$ \tabularnewline
\hline
\end{tabular}\end{center}

\caption{Sensitivity of the PNC asymmetry, $A_{L}\,(\times10^{7})$, to the
effect of a specific correction of the isoscalar PNC $\rho NN$ vertex.
Results are presented for different values of the parameter $\Lambda'$,
introduced in Eqs.~(\ref{eq:fp-km}, \ref{eq:fm-km}). The bare-meson
exchange is adopted for the other components of the PNC potential.
Set S1 is used for the strong parameters, and DDH {}``best-guess''
values for the weak coupling constants.}

\label{tab:vertexcorrection}
\end{table}

Looking at the results given in Tab.~\ref{tab:vertexcorrection},
it is seen that the magnitude of the asymmetry increases when the
parameter $\Lambda'$ becomes smaller. This can be understood from
the behavior of the potential. In Fig.~\ref{fig:ff-comp}, we plot
the modified Yukawa potential multiplied by $r^{2}$ for the PNC chiral-soliton
form factor (right). With a smaller $\Lambda'$ value, the position
of the peak is shifted to larger $r$, and the curve becomes broader.
This behavior leads to a large enhancement in the range $0.5\leq r\leq2$
fm. In the result with the $2\pi+N^{*}$ corrections, we discussed
that a large portion of the difference in the asymmetry is expected
to be originated from the different behavior of the potential in the
range $0.4<r<2$ fm. The behavior of $f_{\rho+}^{\mathrm{KM}}(r)$
and $f_{\rho-}^{\mathrm{KM}}(r)$ in Fig.~\ref{fig:potbare-kmr2}
supports this conjecture; and though these functions even change sign
at $r\leq0.2$ fm, enhanced results are still found. A smaller $\Lambda'$
value gives rise to a more enhanced potential in the range $0.4<r<2$
fm and consequently, this gives a larger magnitude of asymmetry. 

Similarly to the effect discussed in the previous section, the enhancement
of the potential in the range $0.5\leq r\leq2$ fm can explain the
enhancement of the asymmetries calculated in this section with respect
to the bare-$\rho$ ones. Again, the enhancements are larger at low
than at high energy (roughly 73\% and 56\% for $\Lambda'=0.771$ GeV)
but the relative difference is more obvious here (17\% instead of
$0\sim3$\%). It is interesting to compare the above results with
the PWBA ones. In this case, the effect of the form factor under consideration
provides a slight suppression at low energy while, at high energy,
it leads to a large enhancement, 80\% for $\Lambda'=0.771$ GeV (a
factor 3 for the $P$ to $D$ states transition at low energy). These
results evidence a striking feature. While the enhancement for the
$P$ to $D$ states transition amplitude at high energy in PWBA is
more or less recovered by the actual DWBA calculation, the appearance
of an enhancement for the $S$ to $P$ states transition amplitude
at low energy in PWBA is much less expected by DWBA. To some extent,
this confirms the conclusion from considering similar longer-range
forces as due to monopole form factors or a non-zero width of the
$\rho$ meson that: contrary to a naive expectation, a longer-range
force does not necessarily imply an enhancement of $P$ to $D$ states
transition amplitudes at high energy over the $S$ to $P$ states
one at low energy. 

Somewhat surprised by this last result, we looked for an explanation.
It turns out that both the $^{1}S_{0}$ and $^{3}P_{0}$ wave functions
entering the $S$ to $P$ transition amplitude are strongly suppressed
at short distances in the AV18 model, which we used to describe the
strong $NN$ interaction. This suppression acts the same way as the
centrifugal barrier favors Born amplitudes with higher orbital angular
momenta. However, in AV18, the suppression for the $^{1}S_{0}$ and
$^{3}P_{0}$ is even stronger than the one for $^{3}P_{2}$ and $^{1}D_{2}$,
so an opposite situation occurs here: a longer-range PNC force enhances
the $S$ to $P$ transition amplitude with respect to the $P$ to
$D$ one. This feature is largely due to the $^{3}P_{0}$ wave function
where the effect of a short-range repulsion extends to medium distances.
Were this wave function similar to the $^{3}P_{2}$ one, quite different
results would have been obtained instead. The effect of a longer-range
interaction would then be more similar to what is expected from considering
the PWBA alone. In Ref.~\cite{csbg-prc02}, other strong potential
models were also considered and no big model dependence was found.
Therefore, it can be expected roughly that the above conclusion applies
for cases other than AV18.

\subsection{Fitting the weak coupling constants}

\label{subsec:fits}

Motivated by the values of PNC coupling constants obtained in an earlier
analysis~\cite{csbg-prc02}, we looked in the present work for possible
effects that could affect its conclusions. We consider in this subsection
the quantitative consequences of these effects on the coupling constants. 

We first notice that none of the cases we considered allows one to
reproduce the central values of measurements by relying on presently
known predictions of PNC meson-nucleon couplings. While the $\rho$
exchange dominates, the $\omega$ exchange must necessarily have a
sign opposite to what is expected, confirming Carlson \textit{et al.}'s
analysis, whose main qualitative features were reminded in the introduction.
A question which remains of particular interest is whether the size
of the $\omega NN$ coupling can be made more consistent with expectations.
Before entering into details, we mention that our own results for
the $\rho NN$ and $\omega NN$ couplings, $h_{\rho}^{pp}=-22.2\times10^{-7}$
and $h_{\omega}^{pp}=5.28\times10^{-7}$, slightly differ from Carlson
\textit{et al.}'s ones when the same set of strong couplings, S4,
is used. The discrepancies can be reasonably understood as due to
minor differences in the inputs. 

We begin with the S1 set of strong couplings, for which a qualitative
understanding of the result has been reminded in the introduction.
A least-$\chi^{2}$ fit gives $h_{\omega}^{pp}=10.5\times10^{-7}$
and $h_{\rho}^{pp}=-25.9\times10^{-7}$. The difference with Carlson
\textit{et al.}'s result for $h_{\omega}^{pp}$ is primarily due to
that one in the strong coupling constant $g_{\omega NN}$. The absence
of difference for $h_{\rho}^{pp}$ is somewhat accidental and results
from the cancellation of different effects involving the tensor coupling,
$\kappa_{V}$, the strong coupling constant, $g_{\rho NN}$, and the
monopole form factor, with some being separately discussed below. 

Looking now at the results for the set S2, it is found that the $\omega NN$
coupling obtained from a least-$\chi^{2}$ fit, $h_{\omega}^{pp}=7.2\times10^{-7}$,
is smaller than in the previous case as expected from the discussion
of results in Sec.~\ref{subsec:2pinum}, showing the favorable character
of an enhanced value of the isovector tensor coupling, $\kappa_{V}$,
for the problem under consideration in this work. A value of the $\rho NN$
coupling smaller in magnitude is also obtained, $h_{\rho}^{pp}=-16.3\times10^{-7}$.
Let us elaborate this point in more detail. Sets S2 and S1 differ
by the values of the tensor couplings, $\kappa_{S}$ and $\kappa_{V}$.
Since $|\kappa_{S}|\ll\kappa_{V}$, Eq.~(\ref{eq:roughpd}) for the
$P-D$ transition amplitude approximately implies $h_{\rho}^{pp}\propto1/\kappa_{V}$.
One therefore expects that the ratio of the fitted values, $h_{\rho}^{pp}(S2)/h_{\rho}^{pp}(S1)$,
be close to the ratio of the tensor couplings, $\kappa_{V}(S1)/\kappa_{V}(S2)$.
The approximate equality of the ratios, 0.63 and 0.61 respectively,
confirms that the $\omega$ contribution to the corresponding amplitude
is small. For the set S1, its contribution to the asymmetry at high
energy amounts to 4\%. More generally, the fitted value of $h_{\rho}^{pp}$
depends on the strong coupling constant as $1/(g_{\rho NN}\,\kappa_{V})$.
This implies that the contribution of the term $h_{\rho}^{pp}\, g_{\rho NN}\,\kappa_{V}$
to the $S-P$ transition amplitude, represented by Eq.~(\ref{eq:roughsp}),
is approximately the same for the sets S1 and S2. The remaining term,
$2\, h_{\rho}^{pp}\, g_{\rho NN}+h_{\omega}^{pp}\, g_{\omega NN}\,(2+\kappa_{S})$,
should be therefore the same too. The value of $h_{\omega}^{pp}$
can be fitted so that this term is the same, regardless of the strong
coupling constant. Considering the case discussed above where $\kappa_{V}$
increases, leading to an increase of $h_{\rho}^{pp}$ (algebraically),
it appears that the first contribution to the remaining term, $2\, h_{\rho}^{pp}\, g_{\rho NN}$,
increases with the consequence that the other term $h_{\omega}^{pp}\, g_{\omega NN}\,(2+\kappa_{S})$
has to decrease. This implies that $h_{\omega}^{pp}$ decreases. The
value so obtained is close to the fit one. As will be shown for the
results for the set S3, $h_{\omega}^{pp}$ is also sensitive to the
value of the strong coupling, $g_{\omega NN}$. 

As argued at the end of Sec.~\ref{sub:strong-parameters}, a partly
improved version of the above results obtained with the set S2 could
be given by the {}``$2\pi+N^{*}$'' ones with the set S1. It is
therefore expected that the corresponding fitted couplings should
tend to evidence the same departures to the S1-set results. The value
obtained in the present case for the $\rho NN$ coupling, $h_{\rho}^{pp}=-21.1\times10^{-7}$,
is half way between the results for the S1 and S2 sets. The expectation
is verified for a part, suggesting that the physics which has led
to introduce an enhanced value of $\kappa_{V}$ is not fully accounted
for. The effect of an approximate treatment of the underlying physics
could have more important consequences for the $\omega NN$ coupling.
This one, given by $h_{\omega}^{pp}=11.7\times10^{-7}$, remains close
to the S1-set result. This feature indirectly indicates that the $2\pi$
correction scales the $\rho$-exchange contribution to the low- and
high-energy asymmetries by the same factor, allowing one to account
for its effect by modifying the $\rho NN$ coupling. It results that
the $\omega NN$ coupling is essentially unchanged in the fit procedure.
The lower value of the $\omega NN$ coupling obtained with the S2
set could therefore be questionable to some extent. 

Pursuing the discussion with the results for the set S3, which was
intended to look at the effect of monopole form factors often introduced
to describe hadronic vertices, it is found that the fitted values
of the couplings, $h_{\omega}^{pp}=14.6\times10^{-7}$ and $h_{\rho}^{pp}=-41.1\times10^{-7}$,
are significantly increased (in size) with respect to the previous
ones. The difference with Carlson \textit{et al.}'s results comes
mainly from different values of the strong couplings and of $\kappa_{V}$.
The most important one for the main purpose of the present paper is
due to the value of $g_{\omega NN}$, almost a factor 2, which explains
a large part of the discrepancy between the fitted values of $h_{\omega}^{pp}$.
There are other significant differences but, due to cancellations,
they have not much effect on the PNC $\omega NN$ coupling. Taking
into account that the product $g_{\rho NN}\, h_{\rho}^{pp}\,\kappa_{V}$
is mainly determined by the high-energy point, the difference in the
value of $\kappa_{V}$ is largely compensated by a change in $h_{\rho}^{pp}$.
Differences between the sets S3 and S4 for other ingredients ($g_{\rho NN}$
and $\kappa_{s}$) have a minor effect. Thus, the comparison of results
for these two sets of strong couplings shows that smaller (and more
reasonable) values could be obtained for the weak couplings by increasing
the size of the strong ones but, while this could be suggested by
the phenomenology of the strong $NN$ interaction, there is no theoretical
justification. 

Finally, as for the effects of the weak vertex form factors, the result
for the $\omega NN$ coupling reflects the discussion of the asymmetries
in Sec.~\ref{subsec:ffnum}. The fitted value tends to increase when
$\Lambda'$ decreases. It is given by $h_{\omega}^{pp}=11.5\times10^{-7}$
and $15.2\times10^{-7}$ at $\Lambda'=1.31$ and 0.771 GeV respectively
(correspondingly, $h_{\rho}^{pp}=-16.8\times10^{-7}$ and $h_{\rho}^{pp}=-14.1\times10^{-7}$).
Strictly speaking, the last values obtained for $h_{\rho}^{pp}$ only
apply to $h_{\rho}^{0}$ as only the isoscalar PNC form factor is
taken into account (the isovector and isotensor vertices are still
taken as point like). However, to a good approximation, it can be
considered that the fit determines an effective $h_{\rho}^{pp}$ coupling
given by $h_{\rho}^{0}+0.7\,(h_{\rho}^{1}+h_{\rho}^{2}/\sqrt{6})$
for $\Lambda'=1.31$ GeV and $h_{\rho}^{0}+0.5\,(h_{\rho}^{1}+h_{\rho}^{2}/\sqrt{6})$
for $\Lambda'=0.771$ GeV. From what is left out, which is not expected
to be large, it is in principle possible to get an extra constraint
on the isovector and isotensor couplings. The statistical significance
of the result is expected to be smaller than the $h_{\omega}^{pp}$
one, however.

\section{Conclusion}

We considered the PNC asymmetry in $\vec{p}\, p$ scattering at the
energies 13.6, 45 and 221 MeV, where experimental data are available.
In a recent analysis~\cite{csbg-prc02}, $\rho NN$ and $\omega NN$
weak coupling constants were fitted to reproduce the experimental
data. Though the resulting values are within the reasonable range
given in Ref.~\cite{ddh80}, the fitted $\omega NN$ coupling constant
is opposite in sign to most of the theoretical estimates. Employing
AV18 as a strong interaction model, we investigated the role of the
effects such as different strong coupling constants, cutoffs in the
regularization of the PNC meson-exchange potential, long-range contributions
to its $\rho$-exchange component and PNC form factors of the isoscalar
$\rho NN$ vertex. 

As expected, the asymmetry is sensitive to the strong coupling constants,
on which it depends linearly. Assuming the DDH {}``best-guess''
values for the weak couplings, it was found that all three experimental
data can not be satisfied simultaneously with any of the strong coupling
sets considered in this work. In one case, low-energy results are
within the experimental errors but the high-energy one is not, and
vice versa in the other. Comparison of the results with and without
monopole form factors shows a significant effect. For the cutoff value
$\Lambda=1.31$ GeV, asymmetries are suppressed by about 30$\sim$40\%.
This strong dependence on the cutoff value qualitatively agrees with
the one shown in a different way in Ref.~\cite{csbg-prc02}. Fitting
the weak couplings to the measurements, the authors found that a decrease
of the cutoff value by a factor 0.8 enhances the fitted values of
$h_{\rho}^{pp}$ and $h_{\omega}^{pp}$ from $-22.3\times10^{-7}$
and $5.17\times10^{-7}$ to $-106.7\times10^{-7}$ and $+14.63\times10^{-7}$
respectively~\cite{csbg-prc02}. The $2\pi$-exchange contribution
to the bare-$\rho$-exchange potential gives a sizable enhancement
at both low and high energies. The ratio of enhancement is, however,
similar but slightly larger at 13.6 MeV than that at 45 and 221 MeV.
Consequently, with the $2\pi$ exchange in the PNC potential, the
asymmetries at low energies exceed the experimental ranges, and that
at 221 MeV is close to or within the error bar. Simply speaking, the
$2\pi$ contribution does not change the high-vs.-low energy trend
found in the case of using the bare-meson-exchange potential. The
results with a specific PNC form factor show its strong influence
too. The larger is the change in the potential, the larger is the
magnitude of the asymmetry regardless of the energy: similar to the
case including the $2\pi$ exchange. Concluding this part of our work
based on the DDH {}``best-guess'' values for the weak couplings,
we did not find any effect that could allow one to simultaneously
describe the measurements of the asymmetries at low and high energies. 

As is well known, predictions for the weak $\rho NN$ and $\omega NN$
couplings are uncertain and can largely vary in some range. One can
thus look for values of these couplings which could fit the above
measurements. The striking feature is that in all cases we considered
the $h_{\omega}^{pp}$ coupling constant has a positive sign, opposite
to the DDH {}``best-guess'' one. This is not therefore a surprise
if the above studies with the DDH {}``best-guess'' values could
not provide a good description of the measurements. The sign agrees
with Carlson \textit{et al.}'s one~\cite{csbg-prc02} but the size,
which assumes some improvements in this work, is generally larger,
making it more difficult for the $h_{\omega}^{pp}$ coupling so obtained
to be accommodated in the expected range. Thus, the discrepancy that
motivated the present work, far to be reduced, is enhanced. 

Interestingly, the possibility that the $\omega NN$ coupling be positive
was considered in the past to explain the ratio of the proton-nucleus
force, determined from PNC effects in some complex nuclei, to the
proton-proton one, determined from PNC effects in $\vec{p}\, p$ scattering
at low energy~\cite{ber98}. It was however discarded due to a low
statistical significance and the absence of theoretical support. With
results from incorporating the high-energy point in the analysis of
$\vec{p}\, p$ scattering (Ref.~\cite{csbg-prc02} and present work),
the above prospect becomes less unlikely. 

We here consider three issues. The first one is that the value of
the fitted $\omega NN$ coupling, its sign in particular, is correct.
This implies that present hadronic estimates are missing important
contributions. With this respect, one should distinguish bare and
dressed couplings that could include rescattering effects (loop corrections)~\cite{zhu-2000}.
Including some phenomenology however, it is not clear how much present
estimates should be corrected for them. The second issue is the possible
existence of large corrections to the PNC single-meson exchange potential~\cite{iqb-prc49}.
This concern has motivated various approaches dealing more directly
with $NN$ scattering amplitudes, in the past~\cite{dan-pl,dm-npa}
and quite recently~\cite{zhu-npa}. Multi-meson exchanges or retardation
effects are known to provide large corrections in the strong-interaction
case and there is no reason it should be different here. Along the
same lines, one could also cite relativistic corrections which, for
vector-meson exchanges, could be important~\cite{dm-prc89b}. The
last issue concerns the experiment, especially at the highest energy
of 221 MeV. The discussion throughout the paper is based on the absence
of error bar. Assuming minor adjustments of the meson-nucleon couplings
to the low-energy points, current predictions for the highest-energy
point may be off by a factor two for the central value, but only by
one and a half standard deviation when the experimental error is accounted
for. On the other hand, the PNC experiments are difficult ones and
a naive interpretation of the error bar does not necessarily give
a good indication of where a more accurate measurement would sit.
A tendency to overestimate the real asymmetries has often been observed.
Whatever the issue, we believe all of them quite exciting because
further studies will be required to determine the right answer. 

\begin{acknowledgments}
The work of CPL was supported by the Dutch Stichting vor Fundamenteel
Onderzoek der Materie (FOM) under program 48 (TRI$\mu$P). The work
of CHH was supported by the Korea Research Foundation (grant no. KRF-2003-070-C00015). 
\end{acknowledgments}

\end{document}